\documentclass[12pt]{article}

\usepackage{slashed}
\usepackage{color, verbatim}
\usepackage{latexsym}
\usepackage{amsmath}
\usepackage{graphicx}

\usepackage{cite}
\usepackage{hyperref}

\usepackage{amssymb}
\usepackage{graphicx}
\usepackage{bm}
\usepackage{hyperref}
\usepackage{adjustbox}
\usepackage{multirow}
   
\makeatletter

\@addtoreset{equation}{section}
\makeatother

\begin{document}
\thispagestyle{empty}
IPPP/18/41

\begin{center}

\begin{center}

\vspace{.5cm}

{\Large\sc Top quark FCNCs in extended Higgs sectors}

\end{center}

\vspace{0.8cm}

\textbf{
Shankha Banerjee, Mikael Chala and Michael Spannowsky}\\

\vspace{1.cm}
{\em {Institute of Particle Physics Phenomenology, Physics Department, Durham University,
Durham DH1 3LE, UK}}

%\end{comment}

\end{center}

\begin{abstract}
The large number of top quarks produced at the LHC and possible future hadron colliders 
allows to study rare decays of this particle. In  many well motivated models of new physics, 
for example in non-minimal composite-Higgs models, the existence of scalar singlets can 
induce new flavor-violating top decays surpassing the Higgs contribution by orders of 
magnitude. We study the discovery prospects of rare top decays within such models and develop 
new search strategies to test these interactions in top pair-produced events at the LHC. We 
demonstrate that scales as large as $10$--$50$ TeV can be probed. Improvements by factors of 
$\sim 1.5$ and $\sim 3$ can be obtained at $\sqrt{s} = 27$ TeV and $\sqrt{s} = 100$ TeV 
colliders respectively.

\end{abstract}

\newpage

\tableofcontents

\newpage

\section{Introduction}

Processes mediated via Flavour Changing Neutral Currents (FCNC) are very rare within the Standard Model (SM) of 
particle physics. Any experimental evidence of such processes will thus serve as a clear signal for new physics. 
These possible rare processes have garnered strong interest in both the theoretical and experimental communities 
and have triggered numerous studies in search for FCNCs, particularly in the top production~\cite{Tait:1996dv, Guasch:2006hf, Plehn:2009it,Blanke:2013zxo, Degrande:2014tta, Goldouzian:2014nha, Backovic:2015rwa} 
and decay channels~\cite{Agashe:2009di,Mele:1998ag,Greljo:2014dka,Azatov:2014lha,Khachatryan:2015att,Botella:2015hoa,
Bardhan:2016txk,Badziak:2017wxn,Khachatryan:2016sib,Gabrielli:2016cut,CMS-PAS-TOP-17-017,Aaboud:2018nyl,Aaboud:2017mfd,Aaboud:2018pob,Papaefstathiou:2017xuv}. 
Looking for such rare processes in the top sector is very lucrative 
because top quarks are copiously produced at high-energy hadron colliders, and therefore a large number of events 
can be expected even if the FCNC top decays are very rare. Furthermore, because of its large mass, the top quark is 
inherently connected to the Electroweak Symmetry Breaking (EWSB) sector where new physics effects are more likely 
to be present. Numerous experimental searches have been carried out in the top FCNC sector. Some of these searches
include $t\rightarrow \gamma c/u$~\cite{Khachatryan:2015att}, $t\rightarrow gc/u$~\cite{Khachatryan:2016sib}, 
$t\rightarrow Zc/u$~\cite{CMS-PAS-TOP-17-017,Aaboud:2018nyl} and $t\rightarrow hc/u$~\cite{Aaboud:2017mfd,Aaboud:2018pob} 
in single top production or top-pair production. The present experimental bounds on these decays are 
$\mathcal{B}(t\rightarrow \gamma c \; (u)) < 1.7 \; (0.13) \times 10^{-3}$~\cite{Khachatryan:2015att}, 
$\mathcal{B}(t\rightarrow gc \; (u)) < 4.1 \; (0.2) \times 10^{-4}$~\cite{Khachatryan:2016sib}, 
$\mathcal{B}(t\rightarrow Z c \; (u)) < 2.4 \; (1.7) \times 10^{-4}$~\cite{CMS-PAS-TOP-17-017,Aaboud:2018nyl} and 
$\mathcal{B}(t\rightarrow h c \; (u)) < 2.2 \; (2.4) \times 10^{-3}$~\cite{Aaboud:2017mfd,Aaboud:2018pob}.
% On top of that, the neutron EDM constraints ensuing from the CKM phase, bound the $t \to h u$ decay rate to be less than $10^{-7}$~\cite{Dedes:2014asa}.
Similar conclusions can also be seen from the following 
references~\cite{Craig:2012vj,Harnik:2012pb,Backovic:2015rwa}.

\textit{A priori}, however, there might be other particles lurking around the Electroweak (EW) scale, into which 
the top quark can possibly decay. Out of the different possibilities, scalar singlets, $S$, constitute a prime 
example at the level of dimension-four interactions. When the singlet mixes with the SM-like Higgs boson, its 
production is strongly constrained owing to the increasingly precise Higgs signal strength 
measurements (central values reaching unity along with decreasing 
uncertainties) in various channels~\cite{Khachatryan:2016vau, Banerjee:2015hoa} and also from measurements of the $W$-boson mass~\cite{Alcaraz:2006mx,Aaltonen:2012bp,D0:2013jba,Robens:2015gla,Lopez-Val:2014jva}. For a relatively 
small mixing parameter of $\sin\theta$ varying between 0.2 and 0.35~\cite{Peskin:2012we, Craig:2014lda}, one can however, have a 
wide range of allowed singlet mass.

Such scalar particles are predicted by some of the best motivated models of new physics, including 
supersymmetric extensions (\textit{e.g.,} the NMSSM~\cite{Ellwanger:2009dp}) and Composite Higgs Models 
(CHM)~\cite{Dimopoulos:1981xc,Kaplan:1983fs,Kaplan:1983sm,Panico:2015jxa}. Moreover, we will see that these scalars can induce FCNCs significantly larger 
than those mediated by the SM-like Higgs boson~\cite{Zhang:2013xya}. The reason for the above is three-fold; \textit{(i)} The top FCNCs 
mediated by a new scalar singlet are generally suppressed by one less power of the heavy physics scale, 
\textit{(ii)} in principle, the scalar singlet can have a larger decay width into cleaner final states, such as 
$\ell^+\ell^-$, $b\overline{b}$ or $\gamma\gamma$ and \textit{(iii)} in broad classes of models (\textit{for example} 
in CHMs), Higgs mediated FCNCs are forbidden in first approximation~\cite{Agashe:2009di}. Altogether, top FCNCs 
mediated by new scalars might be well within the reach of the LHC.	

Presently, there are no direct limits on $t \to q S$. However, one can have strong constraints from $D^0-\bar{D}^0$ oscillations~\cite{Bona:2007vi,Harnik:2012pb,Agashe:2013hma} which always come about as a product of two $S$ Yukawas, $Y_{ct}$ and $Y_{ut}$ (and also $Y_{uc}$). In order to circumvent these constraints, one can always fall back upon scenarios where $Y_{ut}$ is negligibly small. We will argue in section~\ref{sec:models}, that the $ut$ FCNCs can be vanishingly small compared to their $ct$ counterpart in explicit models. 

In this work, we scrutinise the reach of the LHC for top FCNCs in top-pair produced events. We consider the 
standard leptonic decay of one of the tops, while the other is assumed to decay into $S c$, with either 
$S\rightarrow b\overline{b}$ or $S\rightarrow\gamma\gamma$ (leptonic decays will be analysed in a later work). 
In principle, the current experimental searches for $t\rightarrow hc$ might be also sensitive to these signals. 
However, these searches are only optimised for a $125$ GeV scalar resonance and not for the whole range of masses, 
in which $S$ can potentially lie. Moreover, the latest experimental strategies (see \textit{e.g.} Ref.~\cite{CMS:2017cck}) 
rely on trained BDTs which make them hard to recast for arbitrary scalar masses. In light of these issues, we 
develop new dedicated analyses tailored for each mass point.

The structure of the paper is as follows. In section~\ref{sec:eft} we outline a model-independent introduction to 
the signals of interest. We follow this up in section~\ref{sec:models}, where we discuss concrete realisations, 
involving both strongly and weakly coupled models of new physics. This allows us to establish well-motivated 
benchmark points (BP). We go on to discuss the analysis for the $b\overline{b}$ channel in section~\ref{sec:bb} and 
in section~\ref{sec:gg}, we present the corresponding analysis for the $S \to \gamma\gamma$ mode for the 14 TeV
LHC machine. We finally conclude in section~\ref{sec:conclusions}, where we also provide an outlook for the high energy colliders by commenting on 
naive estimations of the reach of hadron colliders at $\sqrt{s} = 27$ TeV and $\sqrt{s} = 100$ TeV.

\section{Effective Lagrangian}
\label{sec:eft}
Let us consider a scenario where the SM Higgs sector is extended by a gauge singlet, $S$, having a mass $m_S$ in the
EW regime. For low energies, we write the relevant Yukawa Lagrangian as follows~\footnote{We note that in the 
absence of effective operators, the interaction of $S$ with the fermions, is negligible. Indeed, such interactions 
only arise at one loop (and only if $S$ is not a pseudo-scalar; otherwise there is an accidental symmetry 
$S\rightarrow -S$, provided $CP$ is conserved). Moreover, these are proportional to the \textit{vev} of $S$, which 
triggers the mixing with the Higgs boson and are therefore severely constrained by current Higgs 
measurements~\cite{Banerjee:2015hoa,Robens:2015gla,Lopez-Val:2014jva}. In addition, the FCNC currents will be 
further suppressed by the GIM mechanism. Thus, we would expect $\mathcal{B}(t\rightarrow Sc)$ to be several orders 
of magnitude smaller than $\mathcal{B}(t\rightarrow hc)$, which in the SM is predicted to be smaller than
$10^{-13}$ ~\cite{Mele:1998ag}.}
\begin{equation}\label{eq:lag}
 \mathcal{L} = -\mathbf{\overline{q_L}}\bigg(\mathbf{Y} + \mathbf{Y'}\frac{|H|^2}{f^2} + \mathbf{\tilde{Y}}\frac{S}{f}\bigg)  \tilde{H} \mathbf{u_R} + \text{h.c.}~,
\end{equation}
where $H = [\phi^+, (h + \phi^0)/\sqrt{2}]^t$, is the SM-like Higgs doublet, $\mathbf{q_L}$ ($\mathbf{u_R}$) denotes 
the left-handed (right-handed) quarks, $\mathbf{Y}, \mathbf{Y'}, \; \textrm{and} \; \mathbf{\tilde{Y}}$ are arbitrary 
flavour matrices, $v\sim 246$ GeV, is the Higgs vacuum expectation value (\textit{vev}) and $f\gtrsim 
\mathcal{O}(\textrm{TeV})$ is the new physics scale. In general, the flavour matrices are not aligned, and thus the
FCNCs can arise in the EW phase. Among various new physics effects, these induce top flavour-violating decays, 
\textit{viz.}, $t\rightarrow h c$ or $t\rightarrow S c$. In general, the latter dominates over the former, because 
\textit{(i)} $t \to h c$ is further suppressed by an additional factor of $1/f$ and \textit{(ii)} in several 
UV-complete models, $\mathbf{Y}$ and $\mathbf{Y'}$ are approximately aligned. Finally, after the EWSB, one obtains
\begin{align}
\label{lag:lag1}
 \mathcal{L} &= -\frac{v}{\sqrt{2}}\bigg[\mathbf{\overline{q_L}}\mathbf{Y}\left(1+\frac{h}{v}\right)\mathbf{u_R} + \frac{S}{f}\mathbf{\overline{q_L}}\mathbf{\tilde{Y}}\mathbf{u_R} + \mathcal{O}\left(\frac{1}{f^2}\right)\bigg]\supset \tilde{g} \frac{m_t}{f} \overline{t_L} S c_R + \text{h.c.},
\end{align}
where $m_t\sim 173$ GeV is the top mass and $g$ is an $\mathcal{O}(1)$ coupling. Such interactions can be 
tested to a high accuracy through rare top decays. Upon using Eq.~\ref{lag:lag1}, one obtains the partial width of $t \to Sc$ as follows
\begin{equation}
 \Gamma(t\rightarrow Sc) = \frac{\tilde{g}^2}{32\pi}\frac{v^2}{f^2}m_t\bigg(1-\frac{m_S^2}{m_t^2}\bigg)^2~,
\end{equation}
and thus for a benchmark point with $\tilde{g}\sim 1$ and $f\sim 1$ TeV, one obtains $\mathcal{B}(t\rightarrow Sc) 
\sim \Gamma(t\rightarrow Sc)/\Gamma_t^{\text{SM}} \sim 0.03$ with $\Gamma_t^{\text{SM}}\sim 1.4$ GeV~\cite{Patrignani:2016xqp}. 
A full exploration of this decay in singly or pair-produced top quarks at colliders, depends also on how $S$ decays 
to SM particles. Motivated by CHMs (as discussed below), we consider a scenario where $S$ couples to the SM fermions, $\psi$, 
via $\frac{c_{\psi} m_\psi}{f} S\bar{\psi}\psi$ and to the photons via $\frac{c_{\gamma}\alpha}{4\pi f} S F_{\mu\nu}\tilde{F}^{\mu\nu}$, 
where $c_{\psi}$ and $c_{\gamma}$ are arbitrary $\mathcal{O}(1)$ couplings and $\alpha$ is the fine-structure constant. Thus, in
the regime, $m_S \gg m_\psi$, one obtains at leading order
\begin{equation}
 \Gamma(S\rightarrow\psi\psi) = \frac{N_c}{8\pi}\frac{c_{\psi}^2 m_\psi^2}{f^2}m_S~ \; \textrm{and} \; \quad \Gamma(S\rightarrow\gamma\gamma) = \frac{c_\gamma^2 \alpha^2}{64 \pi^3 f^2}m_S^3~.
\end{equation}
Thus, one finds the relation $\mathcal{B}(S\rightarrow\gamma\gamma)/\mathcal{B}(S\rightarrow \overline{\psi}\psi) \sim 
\frac{\alpha^2}{\pi^2} (m_S/m_\psi)^2$. The suppression factor driven by the small electromagnetic coupling can 
thus be partially compensated upon scaling with the free parameter $m_S$.  
$\mathcal{B}(S\to \gamma\gamma)$ can be significantly larger than 
the $\mathcal{B}(h\to \gamma\gamma) \sim 2 \times 10^{-3}$, with the precise value being 
model dependent.
\section{Explicit models}
\label{sec:models}
The Lagrangian in Eq.~\ref{eq:lag} appears naturally in several UV-complete models, for example in CHMs. 
In these classes of models, $H$ and $S$ are pseudo Nambu-Goldstone Bosons (pNGBs) arising in a new global symmetry 
breaking $\mathcal{G}/\mathcal{H}$ at a scale $\sim f$. A prime example is the CHM based on the coset 
$SO(6)/SO(5)$~\cite{Gripaios:2009pe}, which is the smallest one that admits four-dimensional UV completion~\cite{Ferretti:2013kya}. 
The generators of this coset can be chosen as
\begin{align}\label{eq:base}
T^{mn}_{ij} &= -\frac{\mathtt{i}}{\sqrt{2}} (\delta^m_i\delta^n_j-\delta^n_i\delta^m_j)~,
 &m<n\in[1,5]~,\\[-1mm]
X^{m6}_{ij} &= -\frac{\mathtt{i}}{\sqrt{2}} (\delta^m_i\delta^6_j-\delta^6_i\delta^m_j)~,
 &m\in[1, 5]~.
\end{align}
Among these, $X^{16}-X^{46}$ expand the coset space of the Higgs doublet, while the broken generator associated to 
$S$ is provided by $X^{56}$. The SM fermions do not couple directly to the (fully composite) Higgs. Instead, the 
latter couples to composite fermionic resonances, which in turn mix with the SM fermions, thus explicitly breaking 
the global symmetry. The Yukawa Lagrangian depends therefore on the quantum numbers of the aforementioned fermionic 
resonances. For concreteness, we will assume that these fields transform in the fundamental representation 
$\mathbf{6}$ of $SO(6)$. The latter can be decomposed as $\mathbf{6} = 1 + 1 + \mathbf{4}$ under the custodial 
symmetry group $SO(4)$. Let us assume that $u_R^i$ is embedded in both singlets, whereas the one for $q_L^i$ is fixed. These are listed as 
\begin{align}
 U_{R_1}^i = (0, 0, 0, 0, \mathtt{i} u_R^i, 0)~, \quad U_{R_2}^i &= (0, 0, 0, 0, 0, u_R^i)~\\
 \mathrm{and}~ \quad Q_L^i &= \frac{1}{\sqrt{2}}(\mathtt{i} b_L^i, b_L^i, \mathtt{i} t_L^i, -t_L^i, 0, 0)~.
\end{align}
Using the corresponding Goldstone matrix
\begin{equation}
 \text{U}=\left[\begin{array}{cccc}
              1_{3\times 3} &  &  & \\
              &1-h^2/(f^2+\Pi) & -hs/(f^2+\Pi) & h/f\\
              &-hs/(f^2+\Pi)  & 1 -S^2/(f^2+\Pi) & S/f\\
              &-h/f & -S/f & \Pi/f^2
             \end{array}\right]~, \Pi = f^2\left(1-\frac{h^2}{f^2}-\frac{S^2}{f^2}\right)^{1/2}~,
\end{equation}
one obtains the Yukawa Lagrangian 
\begin{align}\nonumber
 L &= -f y^{(1)}_{ij}\overline{(\text{U}^T~ Q_L^i)_6}~(\text{U}^T~ U_{R_1}^j)_6 -f y^{(2)}_{ij}\overline{(\text{U}^T~ Q_L^i)_6}~(\text{U}^T~ U_{R_2}^j)_6 +\text{h.c.} \\
 &=-\frac{1}{\sqrt{2}}\overline{q_L^i} h t_R^j\left[ y_{ij}^{(2)}\left(1-\frac{h^2}{f^2}-\frac{S^2}{f^2}\right)^{1/2} + \mathtt{i} y_{ij}^{(1)} \frac{S}{f}\right] + \text{h.c.}~,
 \end{align}
 which, to leading order, reads
 \begin{equation}
 L = -\frac{1}{\sqrt{2}} \overline{q_L^i} h t_R^j \left[y_{ij}^{(2)} -y_{ij}^{(2)} \frac{h^2}{2f^2} + \mathtt{i} y_{ij}^{(1)}\frac{S}{f}+\cdots\right]+\text{h.c.}
\end{equation}
Hence, we obtain $\mathbf{Y}_{ij} = - \mathbf{Y'} = y^{(2)}_{ij}$ and $\mathbf{\tilde{Y}}_{ij} = \mathtt{i} y_{ij}^{(1)}$. 
Thus, to leading order, scalar mediated FCNCs are only driven by $S$, provided that $y_{ij}^{(1)}$ and $y_{ij}^{(2)}$ are not aligned. Even in that case, the FCNCs would still 
arise in the presence of higher-dimensional operators, and then undergo suppression by a factor of $1/g_*^2$ (with 
$g_*$ being a strong coupling) just as in the minimal CHM~\cite{Agashe:2009di,Panico:2015jxa}. Similar results hold 
for other representations, with the exception of those that respect a $S\rightarrow -S$ parity and those for which 
the shift symmetry of $S$ remains unbroken, examples being $q_L^i$ in the $\mathbf{6}$, $t_R^i$ in the 
$\mathbf{15}$~\cite{Chala:2018qdf}.

We note that although $\mathbf{\tilde{Y}}_{ij}$ is in principle arbitrary, one can easily expect sizeable 
top-charm couplings and still have small top-up FCNCs. Indeed, despite being not directly measurable, in 
common viable \textit{ans\"{a}tze}, $\mathbf{Y}$ is hierarchical and nearly block-diagonal, with the maximal 
mixing occurring in the top-charm sector~\cite{Branco:1999tw,Roberts:2001zy}. It can therefore be diagonalised 
as $\mathbf{Y}\rightarrow \mathbf{L}^\dagger \mathbf{Y}\mathbf{R}$, with $\mathbf{L}$ and $\mathbf{R}$ being
block-diagonal as well. Moreover, in CHMs the aforementioned hierarchy reflects the fact that heavier fermions 
couple stronger to the composite sector, so not only to $H$ but also to $S$. One can then easily expect 
 a similar block-diagonal structure for $\mathbf{\tilde{Y}}$. 
As a consequence, $\mathbf{L}\mathbf{\tilde{Y}}\mathbf{R}$ is also block-diagonal with only the top-charm 
mixing.	
Hence, we concentrate on the $t\rightarrow Sc$ decay channel. However, because we will not use any explicit 
$c$-tagging in our analyses, our results can easily be translated for the $t\rightarrow Su$ mode.

We define three Benchmark Points (BP), each including $m_S = 20, 50, 80, 100, 120$ and $150$ GeV, as follows
\begin{align}\nonumber
 \text{BP 1}: \quad \tilde{g} = 1.0~, \quad f =~\, 2~\text{TeV}\quad\Longrightarrow\quad\mathcal{B}(t\rightarrow Sc)\sim 10^{-3}-10^{-2}~;\\\nonumber
 \text{BP 2}: \quad \tilde{g} = 1.0~, \quad f = 10~\text{TeV}\quad\Longrightarrow\quad\mathcal{B}(t\rightarrow Sc)\sim 10^{-4}-10^{-3}~;\\
 \text{BP 3}: \quad \tilde{g} = 0.1~, \quad f =~\, 2~\text{TeV}\quad\Longrightarrow\quad\mathcal{B}(t\rightarrow Sc)\sim 10^{-5}-10^{-4}~; 
\end{align}
We note that, although being \textit{a priori} relatively light, values of $m_S < m_h/2\sim 62.5$ GeV are not 
necessarily excluded by the LHC constraints on the Higgs width~\cite{Khachatryan:2016ctc}. Actually, the latter 
translates into an upper bound $\Gamma(h\rightarrow S S) \lesssim 10$ MeV. Given that for a quartic coupling 
$\lambda_{HS} S^2 |H|^2$ one obtains $\Gamma(h\rightarrow S S)\sim \lambda_{HS}^2/(32\pi) \times v^2/m_h$, we can 
evade the aforementioned bound provided $\lambda_{HS} < 0.05$. Similarly, derivative interactions $\sim h S 
\partial h \partial S/f^2$, unavoidable in CHMs, contribute to the Higgs width with an effective $\lambda_{HS}\sim 
4 m_h^2/f^2\lesssim 0.05$ for a scale, $f \gtrsim 1.2$ TeV.
\begin{table}[t]
\centering
\begin{tabular}{|c|c|c|c|}
\hline
Field & Relevant Lagrangian & Diagram & $\mathbf{\tilde{Y}}_{ij}/f^2$ \\ \hline
%%%%%%%%%%%%%%%%%%%%%%%%%%%%%%%%%%%%%%%%%%%%%%%%
& & &\\[-0.4cm]
$Q = (1, 2)_{1/6}$ & $L_Q = -m_Q\overline{Q}Q + (\alpha_i^Q\overline{Q} S q^i_L$ & \raisebox{-1cm}{\includegraphics[width=0.2\textwidth]{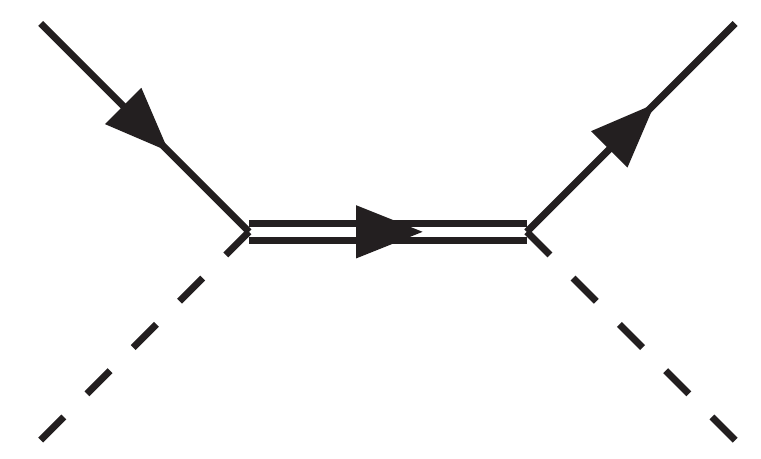}}  &  \raisebox{-0.2cm}{$\dfrac{\alpha_i^Q\tilde{\alpha}_j^Q}{m_Q}$}\\ [-0.6cm]
                   & $+ \tilde{\alpha}_j^Q\overline{Q}\tilde{H}u_R^j+\text{h.c.})$     & &\\
& & &\\[-0.4cm]
$U = (1, 1)_{2/3}$ & $L_U = -m_U\overline{U}U + ( \alpha_i^U\overline{U}Hq_L^i$ & \raisebox{-1cm}{\includegraphics[width=0.2\textwidth]{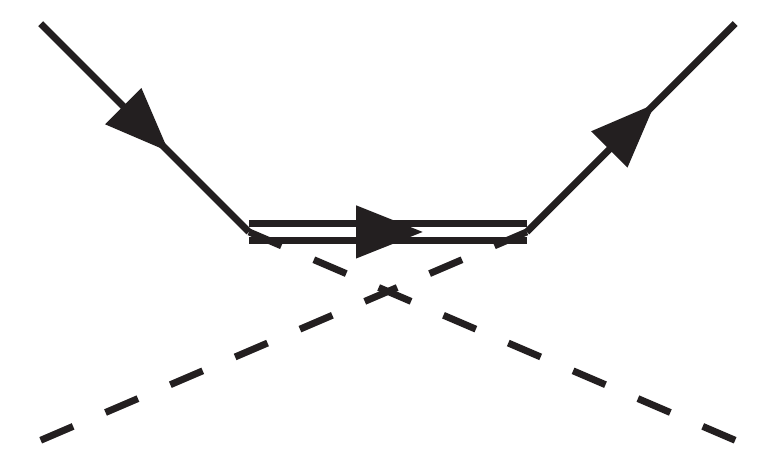}} & \raisebox{-0.2cm}{$\dfrac{\alpha_i^U\tilde{\alpha}_j^U}{m_U}$} \\ [-0.6cm]
                   & $+\tilde{\alpha}_j^U\overline{U} S u_R^j + \text{h.c.})$       & &\\
& & &\\[-0.4cm]
$\Phi = (1, 2)_{1/2}$ & $L_\Phi = -\frac{1}{2}m_\Phi^2\Phi^2 + (\alpha_{ij}^\Phi\overline{q_L^i}\tilde{\Phi}u_R^j$ & \raisebox{-1.cm}{\includegraphics[width=0.2\textwidth]{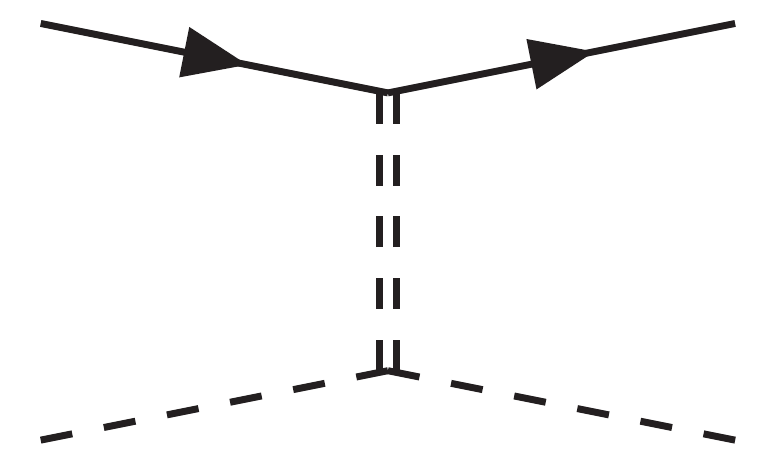}} & \raisebox{-0.2cm}{$\dfrac{\alpha_{ij}^\Phi \kappa}{m_\Phi^2}$} \\[-0.6cm]
& $+\kappa S\Phi^\dagger H + \text{h.c.})$ & &\\[0.3cm]\hline
%
%%%%%%%%%%%%%%%%%%%%%%%%%%%%%%%%%%%%%%%%%%%%%%%%
%
\end{tabular}
\caption{\it Single field extensions of the SM supplemented with $S$ that induce the FCNC of interest at low energy at tree level. The numbers in parenthesis and the subscript denote the $SU(3)_c$ and $SU(2)_L$ representations and the hypercharge, respectively. From the top left and clockwise, the different diagram legs represent $q_L^i$, $t_R^j$, $H$ and $S$, respectively.}
\label{tab:weak}
\end{table}

If we consider only weakly-coupled extensions of the SM$+S$, the Lagrangian in Eq.~\ref{eq:lag} can also be induced 
\textit{at tree level} by the fields listed in Tab.~\ref{tab:weak}. In particular, this means that the NMSSM~\cite{Ellwanger:2009dp}, which extends the SM scalar sector with an additional $SU(2)_L$ doublet with $Y=1/2$ 
(required by SUSY itself) as well as with a singlet (in order to avoid the $\mu$-problem~\cite{Kim:1983dt}), fits 
naturally into the targets of our analysis. 

\section{LHC prospects for $t\rightarrow Sc, S\rightarrow b\overline{b}$}
\label{sec:bb}
%%%%%%%%%%%%%%%%%%%%%%%%%%%%%%%%%%%%%%%

In this section, we focus on the scenario where the scalar singlet decays to a pair of $b$-quarks, yielding a final state comprised
of at least four jets, three of them required to be $b$-tagged and exactly one isolated lepton. As described above, we quantify our results in
terms of six benchmark masses, \textit{viz.}, $m_S = 20, 50, 80, 100, 120 \; \textrm{and} \; 150$ GeV. Our ultimate goal
in this section is to derive an upper bound on $\mathcal{B}(t \to S c, S \to b \bar{b})$ at 95~\% Confidence Level (CL).

We have fixed the $b$-tagging efficiency to 70\%. The $c \; (\ell) \to b$ mistag rate has been taken as 10\% (1\%). 
The most dominant real background ensues from semi-leptonic $t\bar{t} b \bar{b}$ production. Besides, the fully 
leptonic channel from the aforementioned production mode also contributes. The major fake backgrounds that we consider 
are the semi-leptonic (and leptonic) $t\bar{t}$ merged up to one extra matrix element parton, the $Wb\bar{b}$ process 
merged up to two extra matrix element partons and $Zb\bar{b}$ also merged up to two extra partons with the $Z$-boson 
decaying leptonically. For the analysis framework, we use {\tt MG5\_aMC@NLO v2.6.0}~\cite{Alwall:2014hca} for generating the 
signal and background samples. We employ the MLM merging scheme~\cite{Mangano:2006rw} 
embedded in this framework, with appropriate parameter choices. We use very loose parton level cuts, \textit{viz.}, $p_T(j) > 15$ GeV, $p_T(b) > 15$ GeV
and $p_T(\ell) > 10$ GeV, as well as $|\eta(j)| < 4$, $|\eta(b)| < 4$ and $|\eta(\ell)| < 3$. Moreover, we require the 
$\Delta R$ separations to be zero for each pair at the generation level. The cross sections of $t\overline{t}$, $Wb\overline{b}$,
$Zb\overline{b}$ and $t\overline{t}b\overline{b}$ 
are multiplied by $K$-factors of $1.6$, $2.3$, $1.25$ and $1.13$, respectively. While the first one can be found in
Ref.~\cite{kfactors1}, the other three have been estimated by computing in \texttt{MG5\_aMC@NLO} at NLO in QCD. We use the \texttt{NNPDF 2.3}~\cite{Ball:2012cx} at leading order. The analyses
in this section and in the next are carried out for the 14 TeV LHC.

\begin{figure}[t]
 \includegraphics[width=0.32\columnwidth]{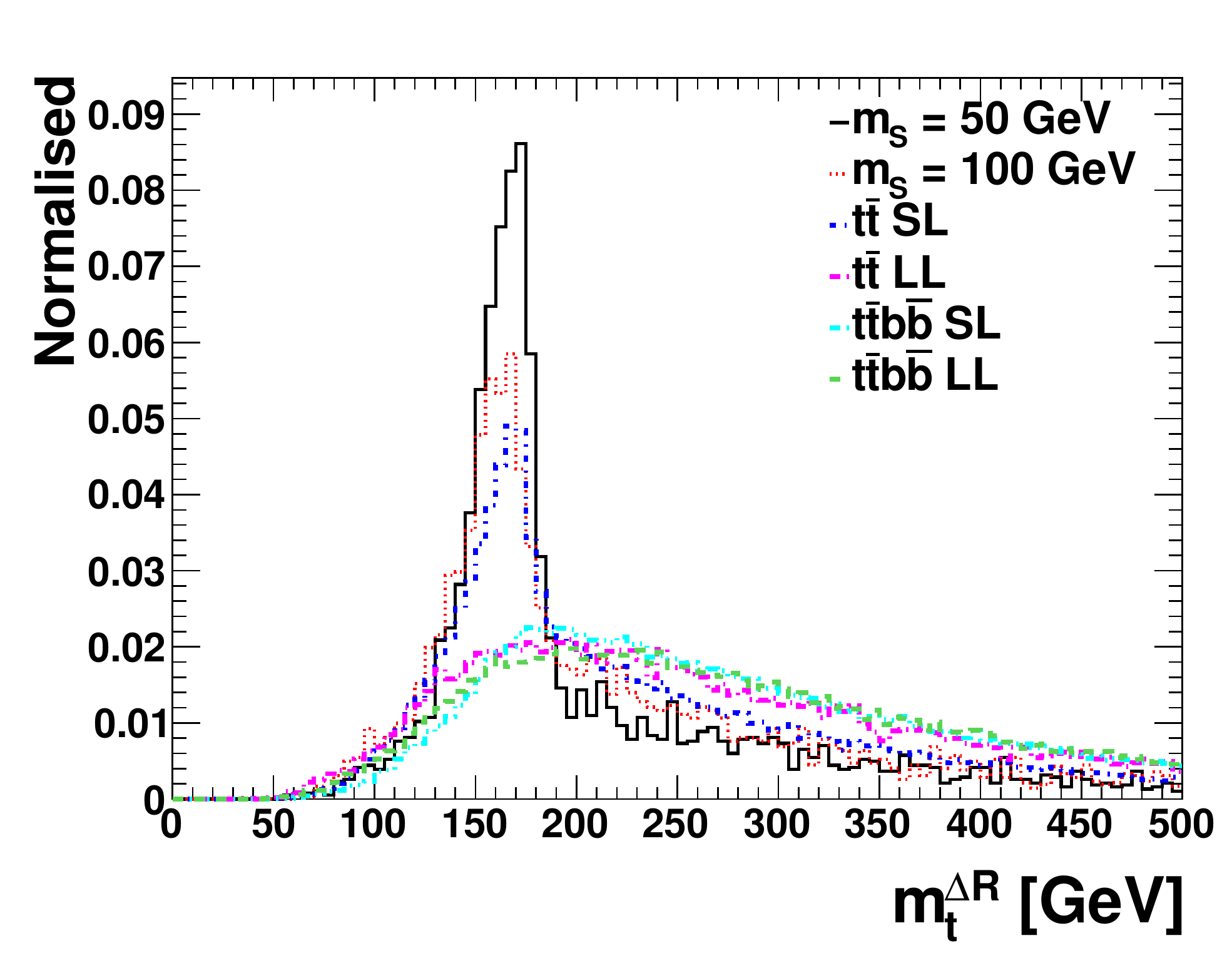}
 \includegraphics[width=0.32\columnwidth]{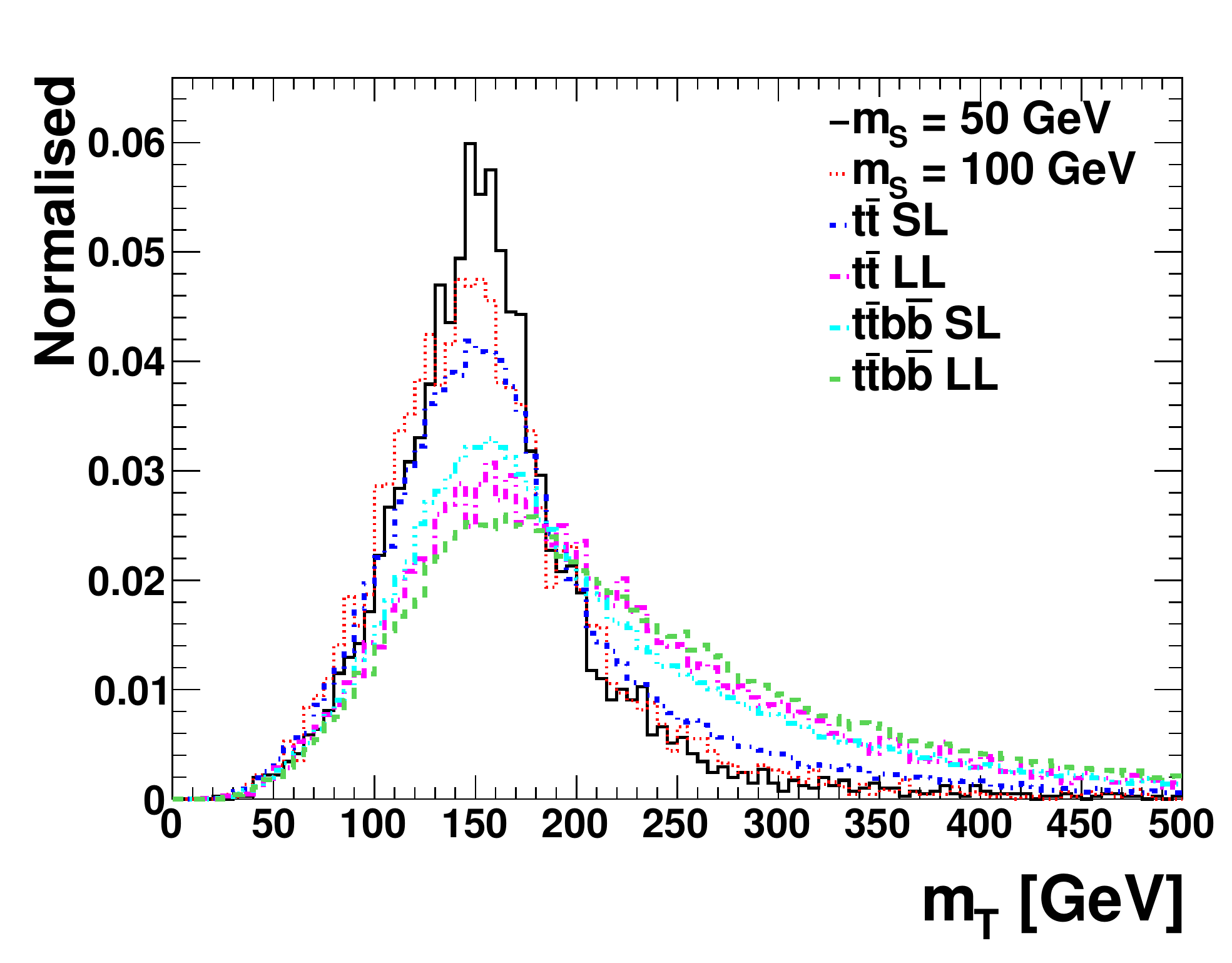}
 \includegraphics[width=0.32\columnwidth]{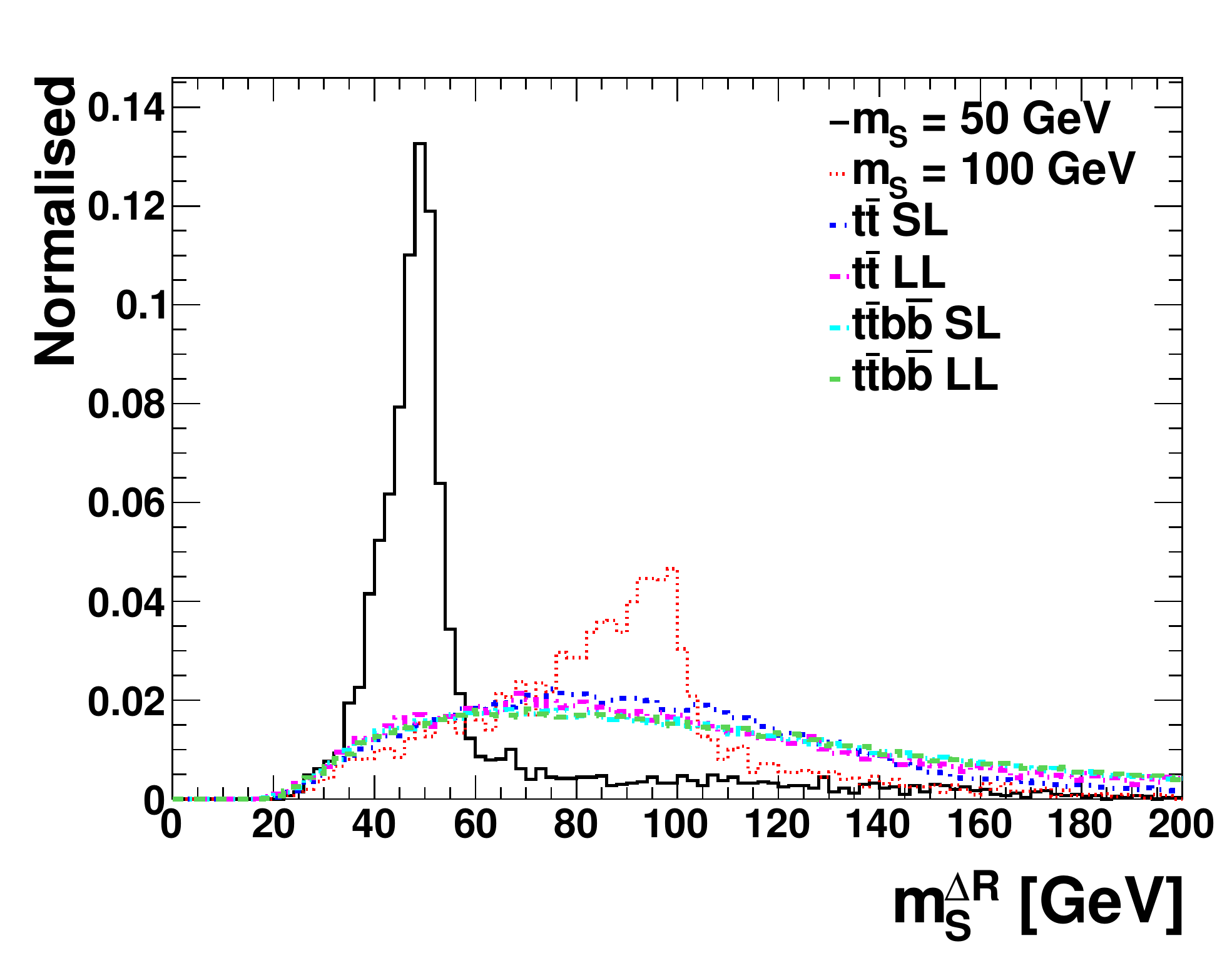}
 \caption{\label{fig:sbb1}\it Left) The reconstructed top mass from the closest $b$-pair. Center) The transverse mass $m_T$. Right) The reconstructed mass of the scalar $S$.}
\end{figure}

%%%%%%%%%%%%%%%%%%%%%%%%%%%%%%%

Furthermore, we shower the samples with \texttt{Pythia 8}~\cite{Sjostrand:2014zea}~\footnote{We did not find any significant differences when comparing with 
another setup using \texttt{MG5\_aMC@NLO v2.1.1} with the showering done with \texttt{Pythia 6}.}. Finally, at the analysis level, 
we construct the jets employing the anti-$k_T$~\cite{Cacciari:2008gp} algorithm with a jet 
parameter $R=0.4$ in the {\tt FastJet}~\cite{Cacciari:2011ma} framework. All the jets are required to have $p_T > 30$ 
GeV and to lie within a pseudorapidity range of $|\eta| < 2.5$. Leptons must have a $p_T > 10$ GeV and $|\eta| < 2.5$. For the isolation, we require that 
the total hadronic activity around the lepton within a cone of $\Delta R = 0.2$ is less than 10~\% of its transverse 
momentum. All the aforementioned selected objects are also required to be separated by $\Delta R > 0.4$.
%
%%%%%%%%%%%%%%%%%%%%%%%%%%%%%%%
\begin{table}[!ht]
\centering
\begin{footnotesize}
\begin{tabular}{| c | c | c | c | c | c | c |}
\hline
Cuts                                              & 20 GeV & 50 GeV & 80 GeV & 100 GeV & 120 GeV & 150 GeV \\
\hline
Basic                                             & 0.014  & 0.050  & 0.051  & 0.056   & 0.063   & 0.063         \\
$|\eta_{(b,\ell,j)}| < 2.5$                       & 0.83   & 0.88   & 0.86   & 0.87    & 0.86    & 0.82         \\
$\Delta R(\text{all pairs}) > 0.4$                & 0.96   & 0.94   & 0.93   & 0.93    & 0.94    & 0.94         \\
$|m_{t}^{\Delta R} - m_t| <$ 50 GeV               & 0.29   & 0.63   & 0.57   & 0.55    & 0.49    & 0.41         \\
$m_T < 200$ GeV                                   & 0.72   & 0.56   & 0.87   & 0.85    & 0.83    & 0.74         \\
\hline
\end{tabular}
\end{footnotesize}
\caption{\label{tab:sbb1} \it Efficiency after each cut for the six signal benchmark points.}
\end{table}
%%%%%%%%%%%%%%%%%%%%%%%%%%%%%%%
%
\begin{table}[!ht]
\centering
\begin{footnotesize}
\begin{tabular}{| c | c | c | c | c | c | c |}
\hline
Cuts                                              & $t\bar{t}$ (SL) & $t\bar{t}$ (LL) & $Wb\bar{b}$ & $Zb\bar{b}$ & $t\bar{t}b\bar{b}$ (SL) & $t\bar{t}b\bar{b}$ (LL) \\
\hline
Basic                                             & 0.0038          & 0.0016          & 0.00032     & 0.00016     & 0.11                    & 0.073                   \\
$|\eta_{(b,\ell,j)}| < 2.5$                       & 0.78            & 0.69            & 0.74        & 0.71        & 0.90                    & 0.85                   \\
$\Delta R(\text{all pairs}) > 0.4$                & 0.95            & 0.94            & 0.95        & 0.95        & 0.96                    & 0.91                   \\
$|m_{t}^{\Delta R} - m_t| <$ 50 GeV               & 0.49           & 0.32            & 0.27        & 0.33        & 0.31                    & 0.28                   \\
$m_T < 200$ GeV                                   & 0.80            & 0.58            & 0.56        & 0.70        & 0.63                    & 0.53                   \\
\hline
\end{tabular}
\end{footnotesize}
\caption{\label{tab:sbb2} \it Efficiency after each cut for the six dominant backgrounds. SL (LL) denotes semi (di)-leptonic decays.}
\end{table}
%%%%%%%%%%%%%%%%%%%%%%%%%%%%%%%

After selecting 
these events, we look for the closest pair (in terms of $\Delta R$ separation) of $b$-tagged jets and reconstruct 
the top-quark mass $m_t^{\Delta R}$ with the additional hardest jet which is not $b$-tagged. We require this 
variable to be within a window of 50 GeV from $m_t$. With the remaining $b$-tagged jet, we construct the 
transverse mass variable $m_T$ and require it to be less than $200$ GeV. We show the distributions of 
$m_t^{\Delta R}$, $m_T$ and the reconstructed scalar mass, $m_S^{\Delta R}$, after the basic cuts 
(which include the aforementioned $p_T$ cuts as well as a requirement for 
3 $b$-tagged jets, at least an additional light jet and one isolated lepton) for two signal benchmark 
points and four dominant backgrounds, in Fig.~\ref{fig:sbb1}. The mass-independent cutflow tables for the six benchmark points and six backgrounds are listed in Tables~\ref{tab:sbb1} 
and~\ref{tab:sbb2} respectively. To optimise each signal region, we impose an additional cut, \textit{viz.}, $0.8\, m_S < m_{S}^{\Delta R} < m_S + 10$ GeV. In Table~\ref{tab:sbb3}, we
list the final efficiencies for each signal region after this additional cut on top of the aforementioned ones.

\begin{figure}[t]
 \includegraphics[width=0.50\columnwidth]{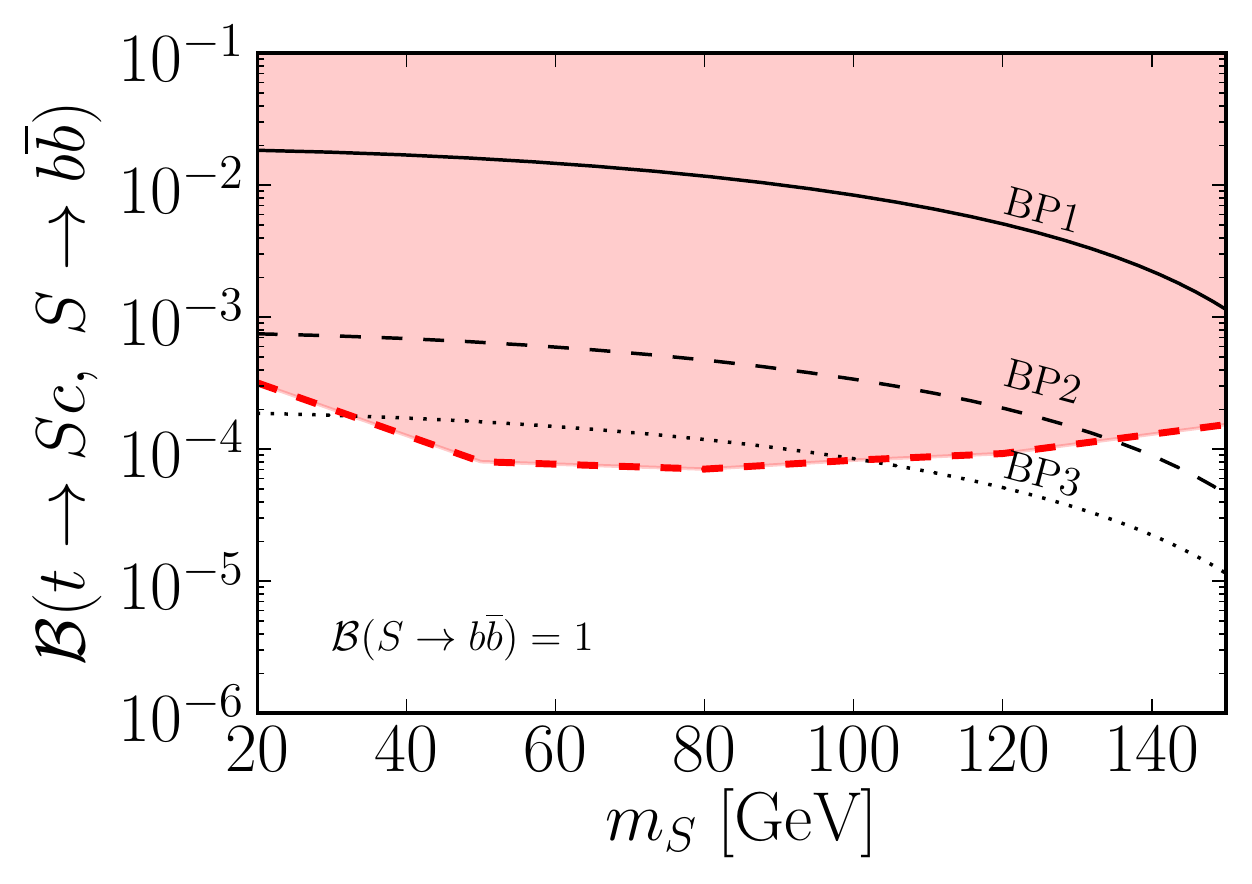}
 \includegraphics[width=0.48\columnwidth]{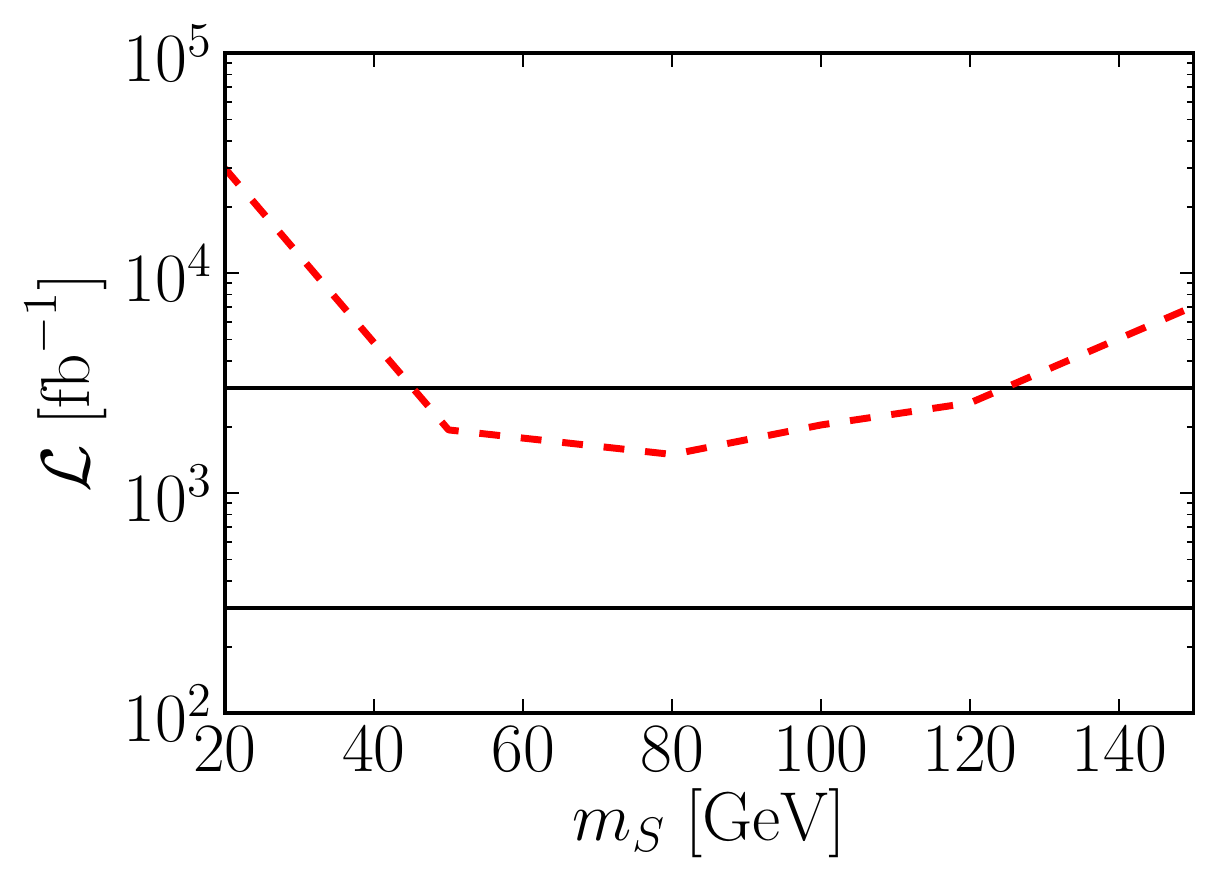}
 \caption{\label{fig:sbb2} \it Left) Branching ratios that can be tested in the $b\overline{b}$ channel. Superimposed are the theoretical expectations in the three BPs. Right) Luminosity required to test $\mathcal{B}(t\rightarrow S c, S\rightarrow b\overline{b}) = 10^{-4}$. Superimposed are $\mathcal{L} = 300$ fb$^{-1}$ and $3000$ fb$^{-1}$.}
\end{figure}

Finally, we show our results in Fig.~\ref{fig:sbb2}. The left plot shows the 95~\% upper 
limit on $BR(t \to S c, S \to b \bar{b})$ and the right plot shows the minimum integrated luminosity to test 
the aforementioned branching ratio to $10^{-4}$ at 95~\% CL.

%%%%%%%%%%%%%%%%%%%%%%%%%%%%%%%
\begin{table}[!ht]
\centering
\begin{footnotesize}
\begin{tabular}{| c | c | c | c | c | c | c | c |}
\hline
$m_S$ [GeV]                                       & Signal  & $t\bar{t}$ (SL) & $t\bar{t}$ (LL) & $Wb\bar{b}$ & $Zb\bar{b}$ & $t\bar{t}b\bar{b}$ (SL) & $t\bar{t}b\bar{b}$ (LL) \\
\hline
20                                                & 8.2     & 0.12          & 0.037          &  0.017       & 0.0094       & 4.0                   & 1.5                   \\
50                                                & 110 & 1.8          & 0.35         & 0.093       & 0.056       & 37                  & 17                  \\
80                                                & 140 & 3.4         & 0.60        & 0.080       & 0.070       & 51                  & 24                  \\
100                                               & 120 &  3.7        & 0.59          & 0.066       & 0.062       & 49                  & 24                  \\
120                                               & 96  & 3.1         & 0.47         & 0.052       & 0.042       & 41                  & 19                  \\
150                                               & 51  & 1.4         & 0.23         & 0.025       & 0.019       & 22                  & 11                  \\
\hline
\end{tabular}
\end{footnotesize}
\caption{\label{tab:sbb3} \it Efficiencies ($\times 10^4$) after the final cut, $0.8\, m_S < m_{S}^{\Delta R} < m_S + 10$ GeV, for each signal benchmark point and for the corresponding backgrounds.  SL (LL) denotes semi (di)-leptonic decays.} 
\end{table}
%%%%%%%%%%%%%%%%%%%%%%%%%%%%%%%
%
%
% %

\section{LHC prospects for $t\rightarrow Sc, S\rightarrow \gamma\gamma$}
\label{sec:gg}
%
% %
We closely follow the previous section in terms of the analysis framework. Here we focus on the scenario in which the scalar
decays to a pair of photons, yielding the final state with at least two jets, with one being $b$-tagged, one
isolated lepton and two isolated photons. Similar to the leptons, we require the photons to have $p_T > 10$ GeV and 
require them to lie within a pseudorapidity range of $|\eta| < 2.5$. We demand the photons to be isolated with the 
exact same criteria as for the leptons as discussed in the section above. The $\Delta R > 0.4$ cuts between pairs of all 
the selected objects are also used for this study. The dominant backgrounds for this channel are the semi-leptonic and
di-leptonic $t\bar{t}h$ processes and the QCD-QED production of $t\bar{t}\gamma\gamma$. The cross section of the former
is scaled by a $K$-factor of $1.68$, what takes into account the NLO corrections to both the production
and the $h$ decay~\cite{kfactors2}. For the second, we use a conservative $K$-factor of $2$. We also include the 
$W\gamma\gamma$ background matched up to two hard jets. However, despite having a cross section of order 
$\mathcal{O}(0.1)$ pb, it becomes irrelevant after imposing all cuts. Consequently, we do not show 
explicit numbers for this process hereafter.

\begin{figure}[t]
 \includegraphics[width=0.49\columnwidth]{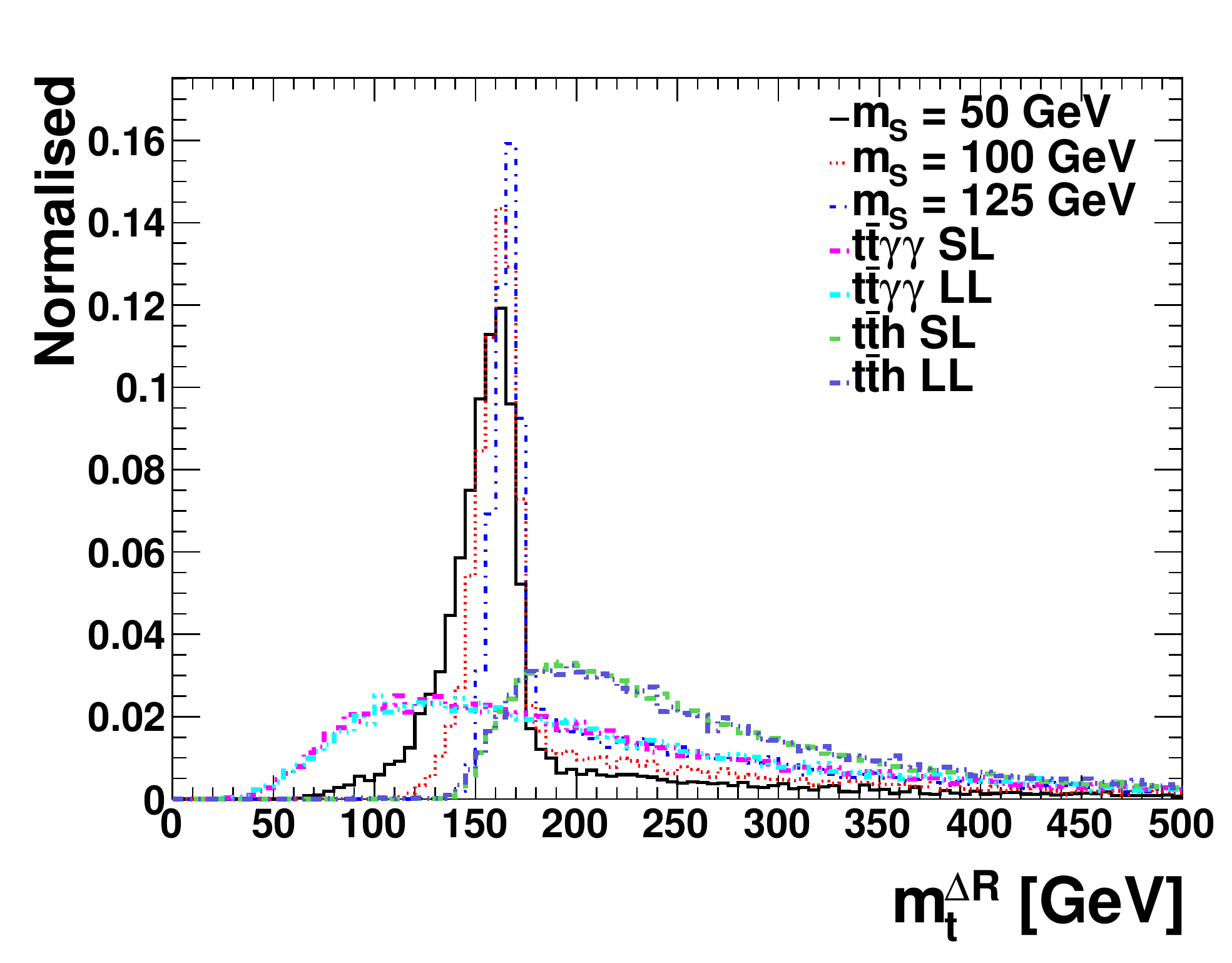}
 \includegraphics[width=0.49\columnwidth]{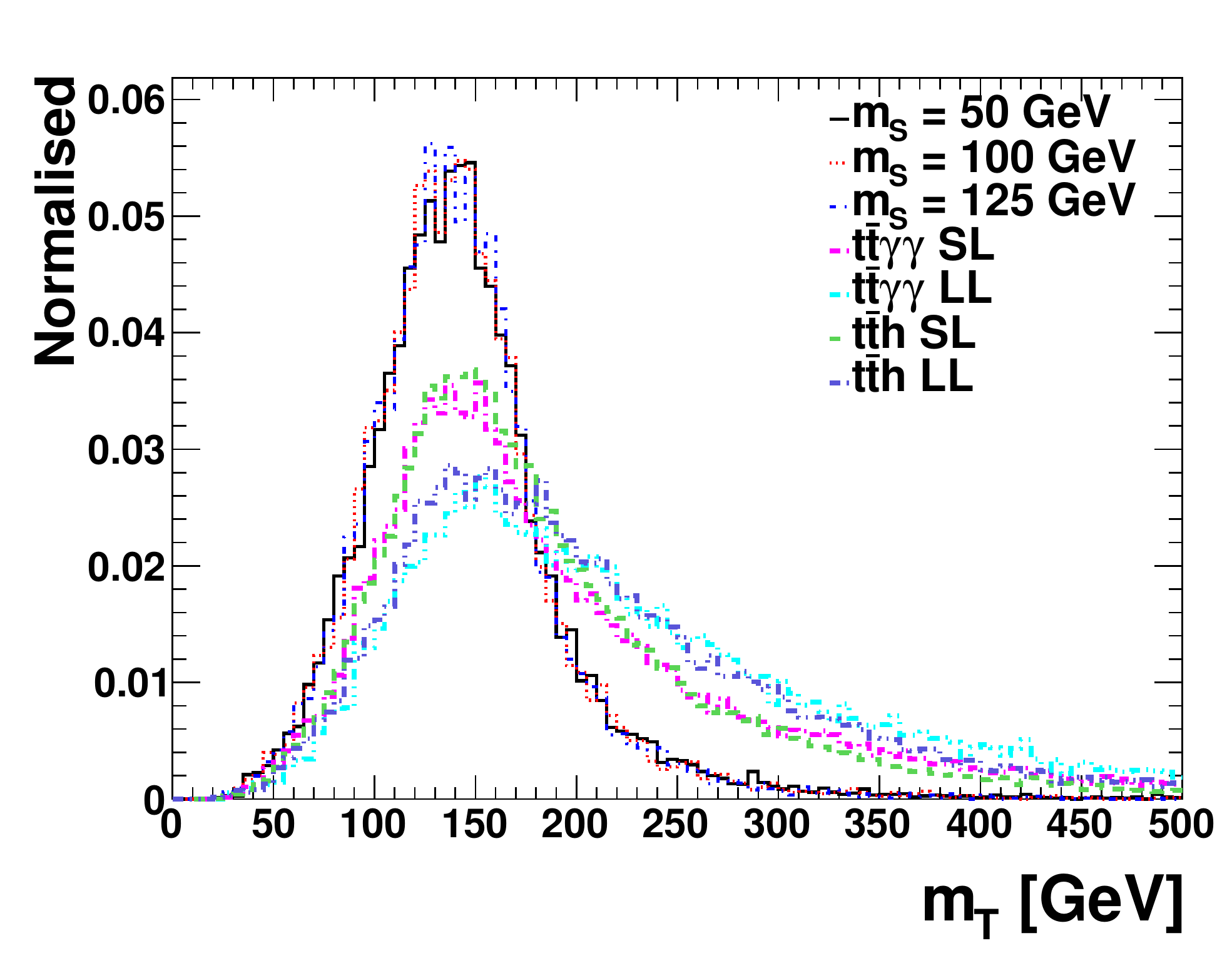}
 \caption{\label{fig:distrgamma}\it Left) The reconstructed top mass from the hardest two photons and the hardest jet. Right) The transverse mass $m_T$. }
\end{figure}

\begin{figure}[t]
 \includegraphics[width=0.50\columnwidth]{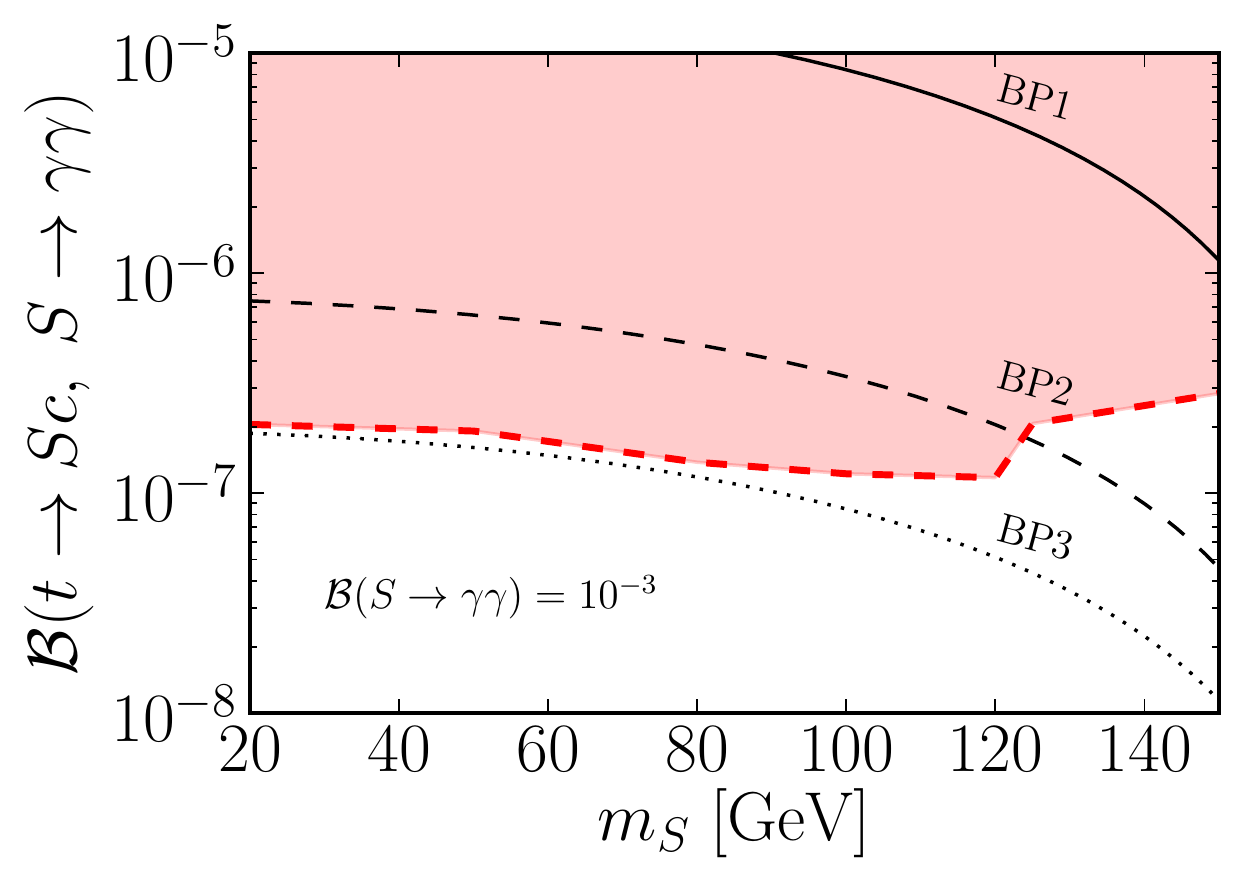}
 \includegraphics[width=0.49\columnwidth]{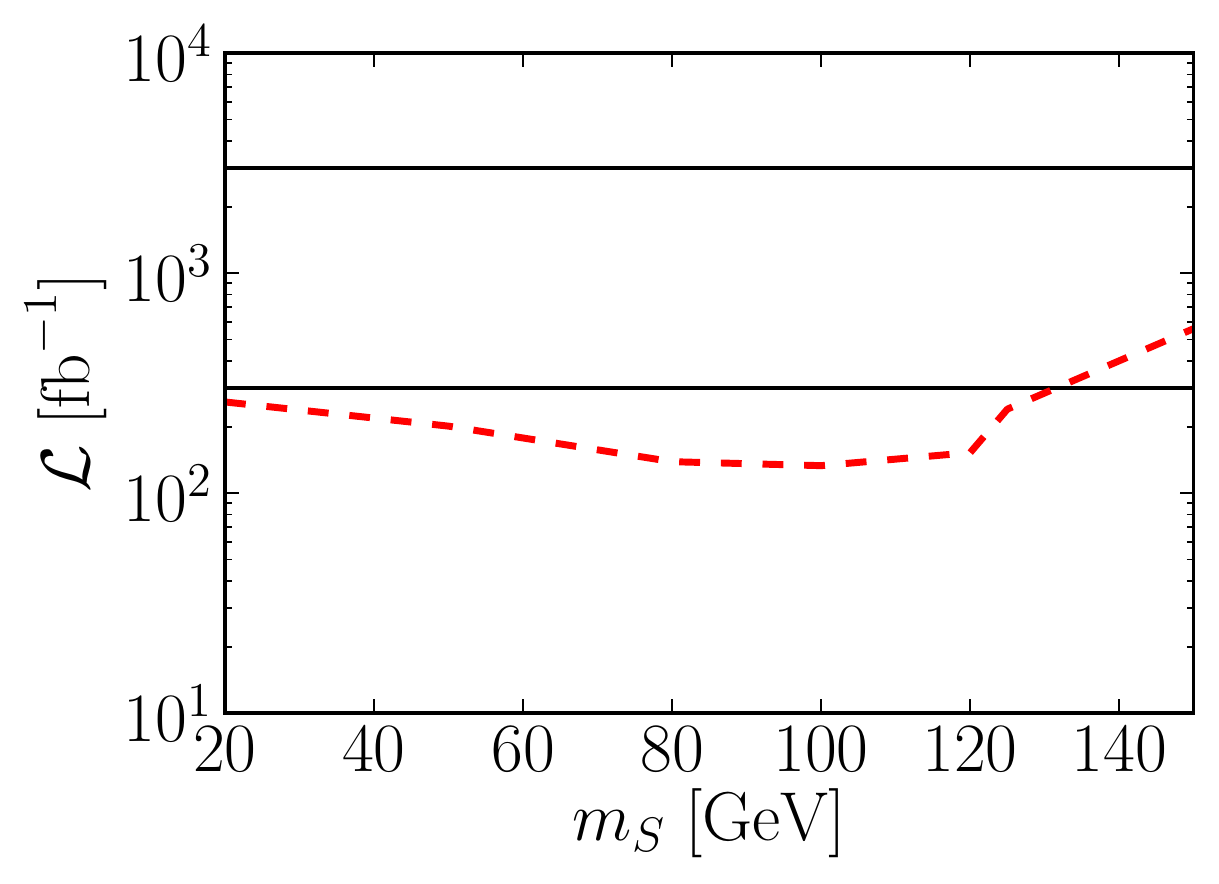}
 \caption{\label{fig:sgg2} \it Left) Branching ratios that can be tested in the $\gamma\gamma$ channel. Superimposed are the theoretical expectations in the three BPs. Right) Luminosity required to test $\mathcal{B}(t\rightarrow St, S\rightarrow \gamma\gamma) = 10^{-6}$. Superimposed are $\mathcal{L} = 300$ fb$^{-1}$ and $3000$ fb$^{-1}$.}
\end{figure}

The selection level cuts up until
the transverse mass are identical to the previous section. However, because of a much sharper diphoton mass resolution,
we demand a very narrow window of 3 GeV around the scalar mass. The shape of the reconstructed top mass distributions
as well as $m_T$ in this case are shown in Fig.~\ref{fig:distrgamma} after the basic cuts (which includes 
objects selected with the $p_T$ requirement as mentioned above along with the selection criteria of exactly 
one $b$-tagged jet, at least one additional jet, one isolated lepton and two or more isolated photons). The cutflows are listed in 
Tables~\ref{tab:sgg1},~\ref{tab:sgg2} and~\ref{tab:sgg3}. The 95\% CL upper limit on the branching ratio
$\mathcal{B}(t \to S c, S \to \gamma \gamma)$ is shown in Fig.~\ref{fig:sgg2} along with the minimum integrated 
luminosity required to probe a branching ratio of $10^{-6}$. In this analysis we have added a new $m_S$ mass
point of $125$ GeV, where the $t\overline{t}h$ background is much larger. The dominance of this latter
process is apparent in the figure.
%%%%%%%%%%%%%%%%%%%%%%%%%%%%%%%
\begin{table}[!ht]
\centering
\begin{footnotesize}
\begin{tabular}{| c | c | c | c | c | c | c | c |}
\hline
Cuts                                                              & 20 GeV & 50 GeV & 80 GeV & 100 GeV & 120 GeV & 125 GeV & 150 GeV \\
\hline
Basic                                                             & 0.18  & 0.18  & 0.18  & 0.17   & 0.16   & 0.15   & 0.12   \\
$|\eta_{(b,\ell,j,\gamma)}| < 2.5$                                & 0.91  & 0.91  & 0.91  & 0.90   & 0.89   & 0.88   & 0.83   \\
$\Delta R(\text{all pairs}) > 0.4$ & 0.62  & 0.91  & 0.91  & 0.90   & 0.88   & 0.88   & 0.87   \\
$|m_t^{reco}-m_t| < 50$ GeV                                      & 0.81  & 0.78  & 0.77  & 0.74   & 0.65   & 0.61   & 0.33   \\
$m_T < 200$ GeV                                                   & 0.93  & 0.92  & 0.92  & 0.93   & 0.93   & 0.94   & 0.94   \\
\hline
\end{tabular}
\end{footnotesize}
\caption{\label{tab:sgg1} \it Efficiency after each cut for the seven signal benchmark points.}\vspace{0.5cm}
\end{table}
%%%%%%%%%%%%%%%%%%%%%%%%%%%%%%%

%%%%%%%%%%%%%%%%%%%%%%%%%%%%%%%
\begin{table}[!ht]
\centering
\begin{footnotesize}
\begin{tabular}{| c | c | c | c | c |}
\hline
Cuts                                                               & $t\bar{t}\gamma\gamma$ (SL) & $t\bar{t}\gamma\gamma$ (LL) & $t\bar{t}h$ (SL) & $t\bar{t}h$ (LL)  \\
\hline
Basic                                                              & 0.18                       & 0.12                       & 0.26            & 0.16             \\
$|\eta_{(b,\ell,j,\gamma)}| < 2.5$                                 & 0.94                       & 0.92                       & 0.94            & 0.89             \\
$\Delta R(\text{all pairs}) > 0.4$  & 0.86                       & 0.72                       & 0.88            & 0.79             \\
$|m_t^{reco}-m_t| < 50$ GeV                                       & 0.37                       & 0.36                       & 0.38            & 0.37             \\
$m_T < 200$ GeV                                                    & 0.65                       & 0.50                       & 0.71            & 0.60             \\
\hline
\end{tabular}
\end{footnotesize}
\caption{\label{tab:sgg2} \it Efficiency after each cut for the four dominant backgrounds. SL (LL) denotes semi (di)-leptonic decays.}\vspace{0.5cm}
\end{table}
%%%%%%%%%%%%%%%%%%%%%%%%%%%%%%%
%

%%%%%%%%%%%%%%%%%%%%%%%%%%%%%%%
\begin{table}[!ht]
\centering
\begin{footnotesize}
\begin{tabular}{| c | c | c | c | c | c |}
\hline
$m_S$ [GeV]                                       & Signal  & $t\bar{t}\gamma\gamma$ (SL) & $t\bar{t}\gamma\gamma$ (LL) & $t\bar{t}h$ (SL) & $t\bar{t}h$ (LL) \\
\hline
20                                                & 760  & 13                       & 5.5                        & 0.15             & 0.20             \\
50                                                & 1100 & 27                       & 9.9                        & 0.40             & 0.25             \\
80                                                & 1000 & 19                       & 6.8                        & 0.45             & 0.35             \\
100                                               & 940  & 13                       & 5.0                        & 0.20             & 0.25             \\
120                                               & 740  & 6.4                        & 3.5                        & 0.25             & 0.35             \\
125                                               & 660  & 5.0                        & 2.6                        & 570           & 240           \\
150                                               & 280  & 2.3                        & 1.1                        & 0.00             & 0.00             \\
\hline
\end{tabular}
\end{footnotesize}
\caption{\label{tab:sgg3} \it Efficiencies ($\times 10^4$) after the final cut, $|m_{\gamma\gamma} - m_S| < 3$ GeV, for each signal benchmark point and for the corresponding backgrounds.  SL (LL) denotes semi (di)-leptonic decays.}
\end{table}
%%%%%%%%%%%%%%%%%%%%%%%%%%%%%%%

\section{Conclusions}
\label{sec:conclusions}
Flavour-violating top decays into scalar singlets, $S$, mediated by new physics at a scale $f\gtrsim 
\mathcal{O}(\textrm{TeV})$, dominate strongly over the ones involving the SM-like Higgs boson. From an 
effective-field theory point of view, the main reason for the dominance of the new scalar is due to the fact that 
the latter proceeds via effective operators of dimension six and is hence suppressed by $1/f^2$, whereas the 
former is already present at dimension five (and hence suppressed only by $1/f$). Moreover, the singlet can
be much lighter than the Higgs, the corresponding top decay being therefore kinematically enhanced. Since such scalar particles 
are predicted in several new physics models, we designed novel analyses dedicated for the upcoming runs of 
the LHC, to search for $t\rightarrow Sc, S\rightarrow b\overline{b}/\gamma\gamma$ in events pertaining to top 
pair production. We restricted our study of $S$ masses varying between $20$ GeV $<m_S<150$ GeV.

In the $S \to b \bar{b}$ channel, the highest reach is obtained for $m_S\sim 80$ GeV, for which we can probe
$\mathcal{B}(t\rightarrow Sc, S\rightarrow b\overline{b}) > 10^{-4}$ at $95$ \% CL with an integrated luminosity 
of $\mathcal{L} = 3$ ab$^{-1}$. The reach is about a factor of $5$ smaller for low masses. This is due to the fact 
that at low masses the two $b$-quarks ensuing from the scalar $S$, do not always form two resolved $b$-jets, and 
hence upon requiring three $b$-tagged jets, we incur a reduction in the efficiency of the signal. However, one 
might consider a fat jet in the framework of a boosted analysis to overcome this difficulty. On the other 
hand, for large masses, the invariant mass of the two $b$-tagged jets closest in $\Delta R$ separation, do not 
always peak at $m_S$.

In the $\gamma\gamma$ channel, the sensitivity of the signal is considerably less dependent on $m_S$, given an 
excellent resolution of the di-photon mass spectrum. We find that a branching ratio, $\mathcal{B}(t\rightarrow Sc, 
S\rightarrow \gamma\gamma) > 10^{-7}$ can be tested at the 95 \% CL with the same integrated luminosity. We note 
that, if $\mathcal{B}(S\rightarrow\gamma\gamma)$ is as large as $\sim 1\%$, then we can indirectly probe new 
physics scales as large as $\sim 50$ TeV. Furthermore, we note that the bound obtained in this channel for 
$m_S\sim m_h$ agrees well with the results obtained for $t\rightarrow hc$ listed in previous 
works~\cite{Papaefstathiou:2017xuv} (which utilise significantly different search strategies and statistical 
approaches). Reference~\cite{Papaefstathiou:2017xuv} also showed that the increase in sensitivity at a 100 TeV 
collider can be roughly estimated by scaling the signal and background cross-sections and the luminosity. In this 
particular channel, the dominant background is $t\overline{t}\gamma\gamma$ for masses of the singlet, $m_S$, well 
separated from $m_h$. It turns out that the increase in cross section in this background at $\sqrt{s} = 27$ TeV 
($100$ TeV) with respect to that at $\sqrt{s} = 14$ TeV is similar to that in the signal, and of order 
$\sim 4$ ($\sim 40$). Thus, assuming an integrated luminosity of $10$ ab$^{-1}$, we expect an increase in 
significance of order $4/\sqrt{4}\times\sqrt{10/3}\sim 3.7$ ($40/\sqrt{40}\times\sqrt{10/3}\sim 11.5$). This 
implies that one can expect up to an order of magnitude improvement in the bound on $\mathcal{B}(t\rightarrow Sc, 
S\rightarrow \gamma \gamma)$. Similar results will hold for the $b\overline{b}$ channel.

\section*{ Acknowledgments}%\\[-3mm]
\noindent
We acknowledge Shilpi Jain, Maria Ramos, Jose Santiago and Jakub Scholtz for useful discussions. MC is supported by the Royal Society under the Newton International Fellowship programme. SB is supported by a Durham Junior Research Fellowship COFUNDed between Durham University and the European Union under grant agreement number 609412.

\noindent

%\clearpage
\bibliographystyle{JHEP}
\bibliography{notes}{}

\end{document}